\documentclass[twocolumn]{article}
\usepackage{amsmath}
\usepackage{amsfonts,amssymb}
\usepackage{bm}
\usepackage{dsfont}
\usepackage[euler]{textgreek}

\usepackage{tikz}
\usetikzlibrary{shapes, arrows, positioning}
\tikzset{
    block/.style={
        draw,
        rectangle,
        minimum height=3em,
        minimum width=6em
    },
    sum/.style={
        draw,
        circle
    }
}

\title{\LARGE \bf
Closed-loop neurostimulation in real-time for the treatment of pathological brain rhythms
}

\author{
    Thomas Wahl%
    \footnote{
        ICube, MLMS, University of Strasbourg; MIMESIS Team, Inria Nancy - Grand Est, Strasbourg, France
    }%
    , Michel Duprez$^*$ and Axel Hutt$^*$
}

\begin{document}

\maketitle
\thispagestyle{empty}
\pagestyle{empty}

%%%%%%%%%%%%%%%%%%%%%%%%%%%%%%%%%%%%%%%%%%%%%%%%%%%%%%%%%%%%%%%%%%%%
\begin{abstract}

\indent 
Mental disorders may exhibit pathological brain rhythms and neurostimulation promises to alleviate of patients' symptoms by modifying these rhythms. Today, most neurostimulation schemes are open-loop, i.e. administer experimental stimulation protocols independent of the patients brain activity which may yield a sub-optimal treatment. We propose a closed-loop feedback control scheme estimating an optimal stimulation based on observed brain activity. The optimal stimulation is chosen according to a user-defined target frequency distribution, which permits frequency tuning of the brain activity in real-time. The mathematical description details the major control elements and applications to biologically realistic simulated brain activity illustrate the scheme's possible power in medical practice. 
\textit{Clinical relevance}— 
The proposed neurostimulation control theme promises to permit the medical personnel to tune a patient's brain activity in real-time.
%This is a brief additional statement on why a this might be of interest to practicing clinicians. Example: This establishes the anesthetic efficacy of 10\% intraosseous injections with epinephrine to positively influence cardiovascular function.
\end{abstract}

%%%%%%%%%%%%%%%%%%%%%%%%%%%%%%%%%%%%%%%%%%%%%%%%%%%%%%%%%%%%%%%%%%%%%%%%%%%%%%%%
\section{INTRODUCTION}
Mental disorders represent clinically significant disturbances in an individual’s cognition, emotional regulation, or behavior. They occur in about a quarter of Europe's adult population~\cite{Wittchen+Jacobi}. In the last decades, research identified characteristic neurophysiological markers that hold the potential to classify mental disorders and reveal their underlying similarities.
For instance, psychotic disorders cause
abnormal thinking and perceptions. Examples of such disorders are bipolar disorder and schizophrenia. Neurophysiological studies have revealed that most mental disorders exhibit certain pathological rhythms that do not occur in healthy patients~\cite{Schulman}. For instance, patients suffering from Parkinson disease exhibit pathological rhythms in the \textbeta-frequency band, i.e. $\sim 12-20$Hz. Electroencephalographic (EEG) studies have shown that psychotic patients have a significantly weak \textalpha-rhythm (oscillations of frequencies between $8$Hz and $12$Hz) and a very strong \textgamma-rhythm (oscillations in the range $[30$Hz$;60$Hz$]$). 

Since such pathological changes in the brain activities power spectrum are characteristic for each mental disorder, a direct and tempting therapy approach attempts to reverse the pathological rhythms and either remove them or render them 'healthy' by neurostimulation. Such an approach has been shown to yield impressive results in Parkinson disease by deep brain stimulation~\cite{desNeiges_etal}, in several disorders by transcranial Direct Current Stimulation~\cite{Tortella} {These techniques} %and in ADHD by neurofeedback~\cite{Enriquez_etal}. The first two examples\cite{desNeiges_etal,Tortella}
are part of a family of open-loop techniques, i.e. the stimulation signal is pre-defined and independent of the current brain's activity.
{Promising results were also obtained} in ADHD by neurofeedback~\cite{Enriquez_etal}{, which along side with \cite{Birbaumer,Ros} comprise} a closed feedback loop and the stimulation is strongly related to neurophysiological signal markers %\MD{separer open loop et closedloop}.
However, the signal feedback is a specific signal feature extracted from the observed brain activity, such as the spectral power in a certain frequency band. Moreover, the feedback signal is typically presented visually on a screen or as auditory information.

The present work combines previous %the current open-loop  and neurofeedback 
techniques by feeding back in real-time a neurostimulation signal computed from observed brain activity. Similar closed-loop neurostimulation methods have been proposed~\cite{Hebb_etal}. For instance, in Deep Brain Stimulation (DBS) pre-defined high-frequency stimulation (typically of $\sim 130$Hz) is applied if the observed brain signals magnitude in the \textbeta-frequency band exceeds a certain threshold. A similar threshold technique in DBS applying pre-defined stimuli is coordinated reset stimulation~\cite{Tass}. Both families of techniques utilize a pre-defined stimulation signal. Conversely, our approach estimates optimally 
the stimulation signal from observed brain signals. The reference in the optimum search for stimulation is a user-defined power spectral density for the observations. To this end, a first method step aims to estimate a dynamical model from observed brain activity yielding an estimate of the brains transfer function. Then, a subsequent user-defined choice of a target transfer function and a closed-loop feedback control permit to stimulate the brain so that the brains activity exhibits the target distribution of spectral density. 

Section \ref{sec:model} is devoted to the presentation of the model estimation method, the closed-loop feedback setup and the brain model under study. %Section~\ref{sec_appl} presents the methods application
In Section~\ref{sec_appl}, we apply our methodology to the brain model and demonstrates, as a perspective, how the medical %personal
 personnel can tune the brains power spectrum by choice.
 We finish with some conclusions and perspectives in Section \ref{sec:concl}. 
 %the The last example  for to reverse such strong have been found to    and may occur in several different  ymptoms of these disorders are diverse and include cognitive Research has been able to identify A large number of mental diseases have been Besides about 27% (equals 82.7 million; 95% CI: 78.5–
%87.1) of the adult EU population, 18–65 of age, is or has been affected by at least one mental disorder in the past 12 months. Taking into account the considerable degree of comorbidity (about one third had more than one disorder), the most frequent disorders are anxiety disorders, depressive, somatoform and substance dependence disorders
%This template provides authors with most of the formatting specifications needed for preparing electronic versions of their papers. All standard paper components have been specified for three reasons: (1) ease of use when formatting individual papers, (2) automatic compliance to electronic requirements that facilitate the concurrent or later production of electronic products, and (3) conformity of style throughout a conference proceedings. Margins, column widths, line spacing, and type styles are built-in; examples of the type styles are provided throughout this document and are identified in italic type, within parentheses, following the example. Some components, such as multi-leveled equations, graphics, and tables are not prescribed, although the various table text styles are provided. The formatter will need to create these components, incorporating the applicable criteria that follow.

\section{Model and closed-loop}\label{sec:model}

We propose a model-based control method for neurostimulation, which aims to increase \textalpha-activity and decrease \textgamma-activity in simulated electroencephalographic data (EEG). This strategy is motivated by pathological EEG observed in psychosis~\cite{Manoach,Reilly}. In the following, we assume that under small stimulation current, the brain response to the current is linear and time-invariant. 
The first key step in this method is identifying the model brain response to neurostimulation. To accomplish this, we utilize the magnitude vector fitting algorithm, as described in \cite{DeTommasi2010}. The second step is the design of a closed-loop control system, including a controller synthesized to produce the desired output frequency distribution, based on the model brain neurostimulation response.
\subsection{Brain model}
The proposed closed-loop feedback scheme is applicable to general models whose evolution can be approximated well by a linear model. Since a large number of successful neurophysiological EEG-models are linear~\cite{NunezBook95}, this constraint does not limit the power of the proposed method. For illustration, we employed a recently developed neural mass model~\cite{Byrne2022} describing the mean neurophysiological currents of two populations $V_{1,2}(t)$ in cortical tissue. Since these currents are known to generate the EEG observed on the scalp of subjects, their sum $y(t) = V_1(t) + V_2(t)$ represents the observed EEG signal. %and extended it by small external random fluctuations.}
%The effectiveness of the method is evaluated by applying it to a brain model based on.
%and comparing the results with previous studies on the subject.
The model details are
\begin{equation}
    \begin{split}
        \tau_1\dot{R_1}(t) &= -\kappa_v^{(1)}R_1(t) + 2R_1(t)V_1(t) + \frac{\gamma_1}{\pi\tau_1} \\
        \tau_1\dot{V_1}(t) &= \eta_1 + V_1(t)^2 - \pi^2\tau_1^2R_1(t)^2 + \kappa_s^{(1)}U_1(t) \\ &+ \xi_1(t) + b_1 u(t) \\
        \tau_2\dot{R_2}(t) &= -\kappa_v^{(2)}R_2(t) + 2R_2(t)V_2(t) + \frac{\gamma_2}{\pi\tau_2} \\
        \tau_2\dot{V_2}(t) &= \eta_2 + V_2(t)^2 - \pi^2\tau_2^2R_2(t)^2 + \kappa_s^{(2)}U_2(t) \\ &+ \xi_2(t) + b_2 u(t),
    \end{split}
    \label{eq:model}
\end{equation}
with $(Q_{1,2}U_{1,2})(t) = R_{1,2}(t)$ and $Q_{1,2} = \left(1 + \frac{1}{\alpha_{1,2}}\frac{\mathrm{d}}{\mathrm{d}t}\right)^2$. This
system is a two population mean-field model, where $R_{1,2}$ represent the mean firing rate of population \#1 and \#2, while $V_{1,2}$ represent their mean potentials. The system is driven by the zero-mean finite size fluctuations $\xi_{1,2}$ with variance $\sigma_{1,2}^2$ and the neurostimulation input current $u$.
%\AH{\bf THOMAS: please explain briefly the meaning of the different variables and parameters}.
All the parameter values and their meaning are given in Table \ref{tab:model}. %The parameters have been chosen to display linearly stable noise driven oscillations in the alpha and gamma frequency ranges.
\begin{table}
    \centering
    \caption{
        Parameter set of the brain model (\ref{eq:model}), see also \cite{Byrne2022}. \\
        All parameters show good accordance to physiological findings~\cite{Koch99}.
    }
    \begin{tabular}{lll}
        \hline\noalign{\smallskip}
        parameter & description & value  \\
        \noalign{\smallskip}\hline\noalign{\smallskip}
        $\tau_1$ & synaptic time constant \#1 &  8 ms \\
        $\tau_2$ & synaptic time constant \#2 & 30 ms \\
        $\alpha_{1,2}$ & synaptic rate constants & 500 Hz \\
        $\kappa_v^{(1)}$ & gap-junction coupling \#1 & 0.3 \\
        $\kappa_v^{(2)}$ & gap-junction coupling \#2 & 0.5 \\
        $\kappa_s^{(1,2)}$ & synaptic coupling & 1.0 \\
        $\eta_{1,2}$ & levels of excitability& 1.0 \\
        $\gamma_{1,2}$ & networks heterogeneity & 0.5 \\
        $N$ & number of neurons & 1000 \\
        $\sigma^2_{1, 2}$ & variance of finite size fluctuations & $0.5/N$ \\
        $b_{1,2}$ & input coupling & 10 \\
        \noalign{\smallskip}\hline
    \end{tabular}
    \label{tab:model}
\end{table}

\subsection{Model estimation}
\label{sec:model_estimation}
The aim of our closed-loop controller is to estimate the brain input response transfer function $G(s),~s\in\mathbb{C}$, which includes the brain dynamics, the neurostimulation device, and the observation device. We employ observed brain activity, such as EEG, to estimate $G(s)$ as accurately as possible. However, this is not a straightforward task since the observed signal is the sum of the resting state activity and the stimulation response.

To address this problem, we first need to extract the stimulation response from the observed signal. To this end, as a first step, an arbitrary time-dependent test input $u(t)$ with time $t$ is applied to the plant, which generates the output
$$y(t) = y_0(t) + y_u(t),$$
where $y_0(t)$ is the brain resting state activity (without stimulation) and $y_u(t)$ is the brain response to the test stimulation $u(t)$ defined by the convolution product $y_u(t):=g(t)\star u(t)$, where $g(t)$ is plant unit impulse response function.

The challenge is that during the stimulation, we can only observe $y$. Therefore, we have to use previous data recordings to predict the resting state activity $y_0$ during the stimulation.%We use the following standard definitions for signal extraction. 
%We assume that $y_0$ is a wide-sense-stationary (WSS) random process, meaning that its mean and variance do not depend on time. According to the Wiener-Khinchin theorem \cite{Khinchin}, the autocorrelation function of a WSS random process has a spectral decomposition given by the power spectrum of that process. 
We assume that the resting state is wide-sense-stationary in time and define the time-dependent signals $\alpha(t)$, $\alpha_0(t)$, $\alpha_u(t)$ as the deviations of $y,~y_0,~y_u$ from their temporal mean. %detrended signals corresponding to $y$, $y_0$ and $y_u$ respectively, i.e. the zero mean signals. 
Then
$$\alpha(t) = \alpha_0(t) + \alpha_u(t),$$
with their corresponding complex-valued Fourier transforms
$$\hat{\alpha}(f) = \hat{\alpha}_0(f) + \hat{\alpha}_u(f)$$
and their square magnitudes
\begin{align*}
    |\hat{\alpha}(f)|^2 %&= |\hat{\alpha}_0(f) + \hat{\alpha}_u(f)|^2 \\
                        %&= (\hat{\alpha}_0(f) + \hat{\alpha}_u(f))(\hat{\alpha}_0(f) + \hat{\alpha}_u(f))^* \\
                        &= |\hat{\alpha}_0(f)|^2 + |\hat{\alpha}_u(f)|^2 + 2\mathrm{Re}(\hat{\alpha}_0(f)\hat{\alpha}_u^*(f)).
\end{align*}
%\MD{plus de détails pour passer aux équations suivantes ou évident ?}
It can be shown that since both $\alpha_0$ and $\alpha_u$ have a zero mean in the time domain, they also have a zero mean in frequency domain. Since $u(t)$ is arbitrary and statistically independent of the resting state activity, applying the Wiener-Khinchin Theorem~\cite{Khinchin}  we can rewrite this equation in terms of spectral densities %while ignoring the $2\mathrm{Re}(\hat{\alpha}_0(f)\hat{\alpha}_u^*(f))$ term \MD{c'est meme nul, non ?}
\begin{equation}
    \begin{split}
        S_{yy}(f) &= S_{y_0y_0}(f) + |\hat{g}(f)|^2S_{uu}(f) \\
        |\hat{g}(f)|^2 &= \frac{S_{yy}(f) - S_{y_0y_0}(f)}{S_{uu}(f)},
    \end{split}
    \label{eq:gmag}
\end{equation}
where $\hat{g}$ is the brain input response transfer function in Fourier space and $S_{\cdot\cdot}(f)$ denotes the power spectral density. The power spectrum $S_{y_0y_0}$ was estimated from the brain activity in the absence of any stimulation, $S_{yy}$ was estimated from the observed brain activity under test input $u(t)$ and $S_{uu}$ was estimated from the arbitrary test input directly. For the power spectrum estimation, we employed Welch's method \cite{Welch}.

\begin{figure}
    \centering
    \includegraphics[width=\linewidth]{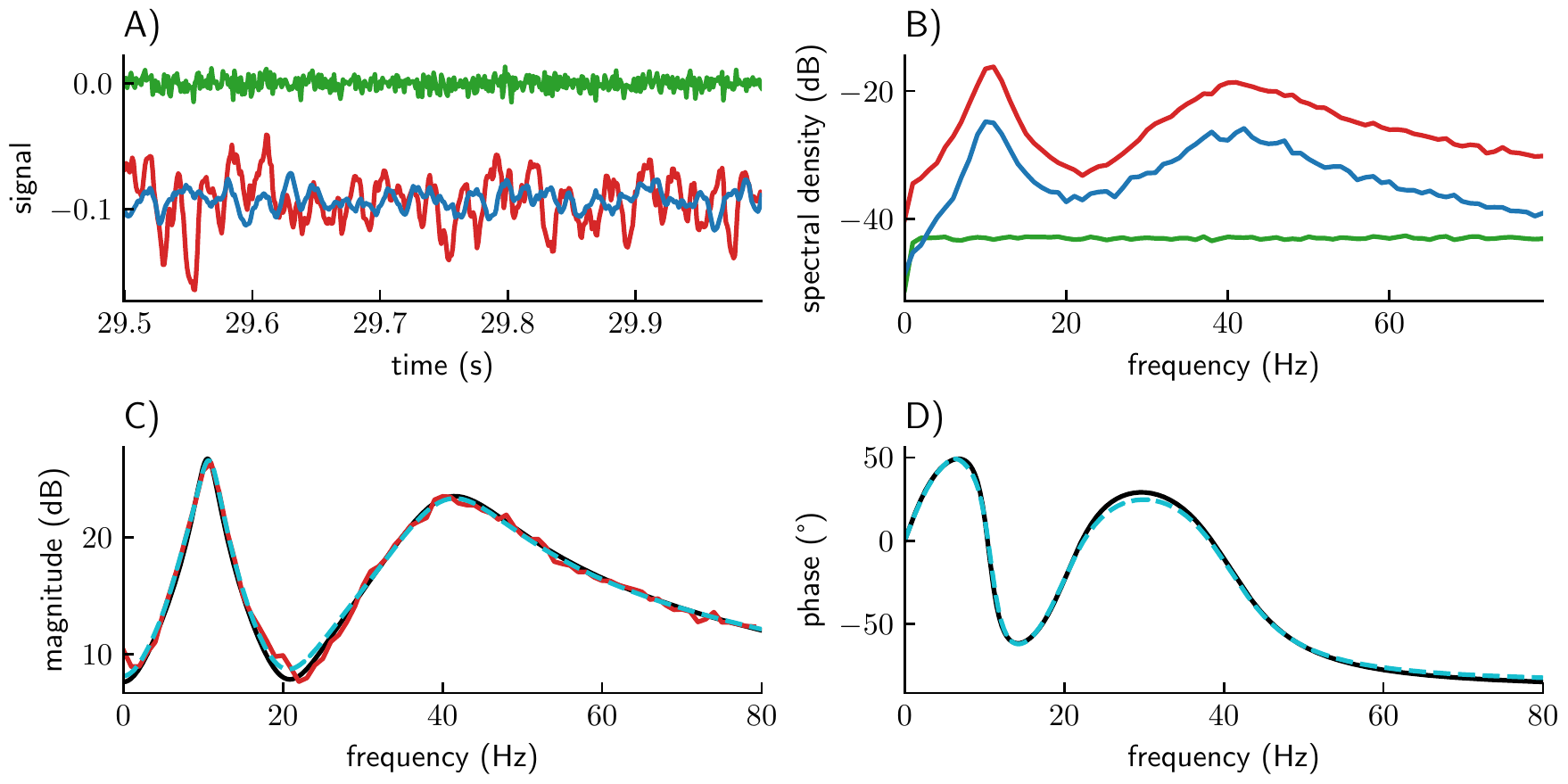}
    \caption{
    {\bf The magnitude vector fitting algorithm performed on open-loop stimulation data accurately reproduces the magnitude and phase shift properties of the brain input response transfer function utilizing the brain model~(\ref{eq:model}).}
    A) Time series of the resting state activity $y_0$ (blue), the resulting brain activity $y$ (red) and the test input current $u$ (green). B) Power spectral densities computed from the respective time series. C) Magnitude of the transfer function $\hat{g}$ of the fitted model (dashed cyan) compared to the magnitude estimated from the spectral densities using \eqref{eq:gmag} (red) and the magnitude of the transfer function of the linearized original model (\ref{eq:model}) (black). D) Phase shift of the transfer function of the fitted model (dashed cyan) compared to the phase shift of the transfer function of the linearized original model (black).
    }
    \label{fig:fit}
\end{figure}

By applying the magnitude vector fitting algorithm \cite{DeTommasi2010} on the computed magnitude data $|\hat{g}(f)|^2$, we estimated $\hat{g}(f)$ and thus obtained a plant model whose dynamics is as close as possible to the brain dynamics. Note that this technique does not require any preliminary knowledge of the underlying brain model, and the model (\ref{eq:model}) is used here only as a black box system for numerical simulations. The accuracy of the estimated model's transfer function $\hat{g}(f)$ is shown in Fig.~\ref{fig:fit}. %\MD{ces simulations estiment le système (4) ? il vaudrait mieux le mettre à cet endroit ou même avant en expliquant que la méthode est générique}

\subsection{Closed-loop neurostimulation}

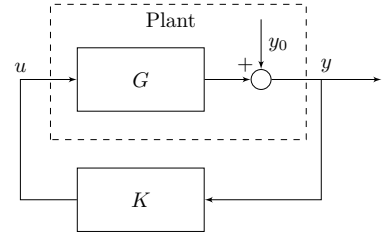
\begin{figure}
    \centering
    \scalebox{.8}{
    \begin{tikzpicture}[auto,>=latex']
        % blocks
        \node [block] at (2, 0) (ctr) {$K$};
        \node [block] at (2, 2) (sys) {$G$};
        \node [sum]   at (4, 2) (sum) {};
        % connections
        \draw (sum) node [above left] {$+$};
        \draw [->] (4, 3) -- node {$y_0$} (sum);
        \draw [->] (sys) -- (sum);
        \draw [->] (5, 2) |- (ctr);
        \draw [->] (ctr) -- (0, 0) |- node {$u$} (sys);
        \draw [->] (sum) -- node {$y$} (6, 2);
        % dashed frames
        \draw [dashed] (.5, 1) rectangle (4.75, 3.25);
        \draw (2.5, 3) node {Plant};
    \end{tikzpicture}
    }
    \caption{
    {\bf Closed-loop neurostimulation circuit.}
    $G$ and $K$ represent the plant input response system and the controller, respectively.    }
    \label{fig:circuit}
\end{figure}

Once the neurostimulation response model is identified, we use it in a closed-loop control scheme to apply the desired modifications to the EEG power spectral density. The control scheme employs a feedback loop, where the EEG signal is fed back to a controller which produces the neurostimulation input signal in real-time based on the identified model, see Fig.~\ref{fig:circuit}.
We define the desired transfer function from $y_0$ to $y$ of the closed-loop system as
\begin{equation}
    1 + H(s),
    \label{eq:closed-loop}
\end{equation}
where $H(s)$ is a weighted double bandpass filter with transfer function
\begin{equation}
    H(s) = c_1\frac{B_1's}{s^2 + B_1's + \omega_1^2} + c_2\frac{B_2's}{s^2 + B_2's + \omega_2^2},
    \label{eq:filter}
\end{equation}
with frequencies $\omega_{1,2} = 2\pi f_{1,2}$, bandwidth $B_{1,2}' = 2\pi B_{1,2}$ and weights $c_{1,2}$. This filter has a frequency band in the \textalpha-range with a positive weighs $c_1$, and a frequency band in the \textgamma-range with a negative weight $c_2$, the parameter values are given in Table \ref{tab:filter}.
\begin{table}
    \centering
    \caption{
        Parameter set of the weighted double bandpass filter, cf. Eq.~(\ref{eq:filter}).
    }
    \begin{tabular}{lll}
        \hline\noalign{\smallskip}
        parameter & description & value  \\
        \noalign{\smallskip}\hline\noalign{\smallskip}
        $f_1$ & \textalpha-band natural frequency  & 10Hz \\
        $B_1$ & \textalpha-band width              & 4Hz \\
        $c_1$ & \textalpha-band weight             & 1.0\\
        $f_2$ & \textgamma-band natural frequency & 40Hz \\
        $B_2$ & \textgamma-band width             & 30Hz \\
        $c_2$ & \textgamma-band weight            & -0.5\\
        \noalign{\smallskip}\hline
    \end{tabular}
    \label{tab:filter}
\end{table}
The motivation for this choice of transfer function is that this filter in the closed-loop transfer function increases the gain of the output signal in the \textalpha-range and decreases the gain of the output signal in the \textgamma-range. From this requirement, and from Fig.~\ref{fig:circuit}, we write an equation for the controller $K(s)$ that produces the desired closed-loop transfer function
%\AH{We propose the controller transfer function} as
\begin{equation}
    K(s) = \frac{H(s)}{\tilde{G}(s)(1 + H(s))},
    \label{eq:controller}
\end{equation}
%\MD{trop rapide ?}\AH{\bf THOMAS: please add one or two sentences on the motivation of this choice}
where $\tilde{G}$ is the transfer function of the fitted model $G$ and $H$ is the transfer function of the filter defining the desired frequency domain modification. Now calling $T(s)$ the closed-loop transfer function from $y_0$ to $y$ and utilizing Fig.~\ref{fig:circuit} and Eq.~(\ref{eq:controller}), we can derive the closed-loop transfer function
\begin{equation}
    \begin{split}
        T(s) &= \frac{1}{1 - G(s)K(s)} \\
             &= \frac{1}{1 - G(s)\frac{H(s)}{\tilde{G}(s)(1 + H(s))}} \\
             &= \frac{1 + H(s)}{1 + H(s)\left(1 - \frac{G(s)}{\tilde{G}(s)}\right)}.
    \end{split}
    \label{eq:tf}
\end{equation}
Under the assumption that $G(s) = \tilde{G}(s)$, this expression simplifies to $1 + H(s)$, which is the desired closed-loop transfer function. Hence, the transfer function of the closed-loop feedback proposed $T(s)$ resembles the reference transfer function $1+H(s)$ for good model approximations $G(s)\approx \tilde{G}(s)$.%\AH{since} $H(s)$ has an \AH{$alpha$-}band with a positive weight and a \AH{\textgamma-}band with a negative weight, the \AH{power} spectral density of $y$ is effectively increased \AH{(decreased)} in the \AH{\textalpha-band} (\AH{\textgamma-band})  compared to \AH{the resting state $y_0$}.
%\AH{\bf THOMAS: please provide the equation for $H(s)$ including the parameters given in Table II.}

\section{Application of closed-loop control}\label{sec_appl}
\begin{figure}
    \centering
    \includegraphics[width=\linewidth]{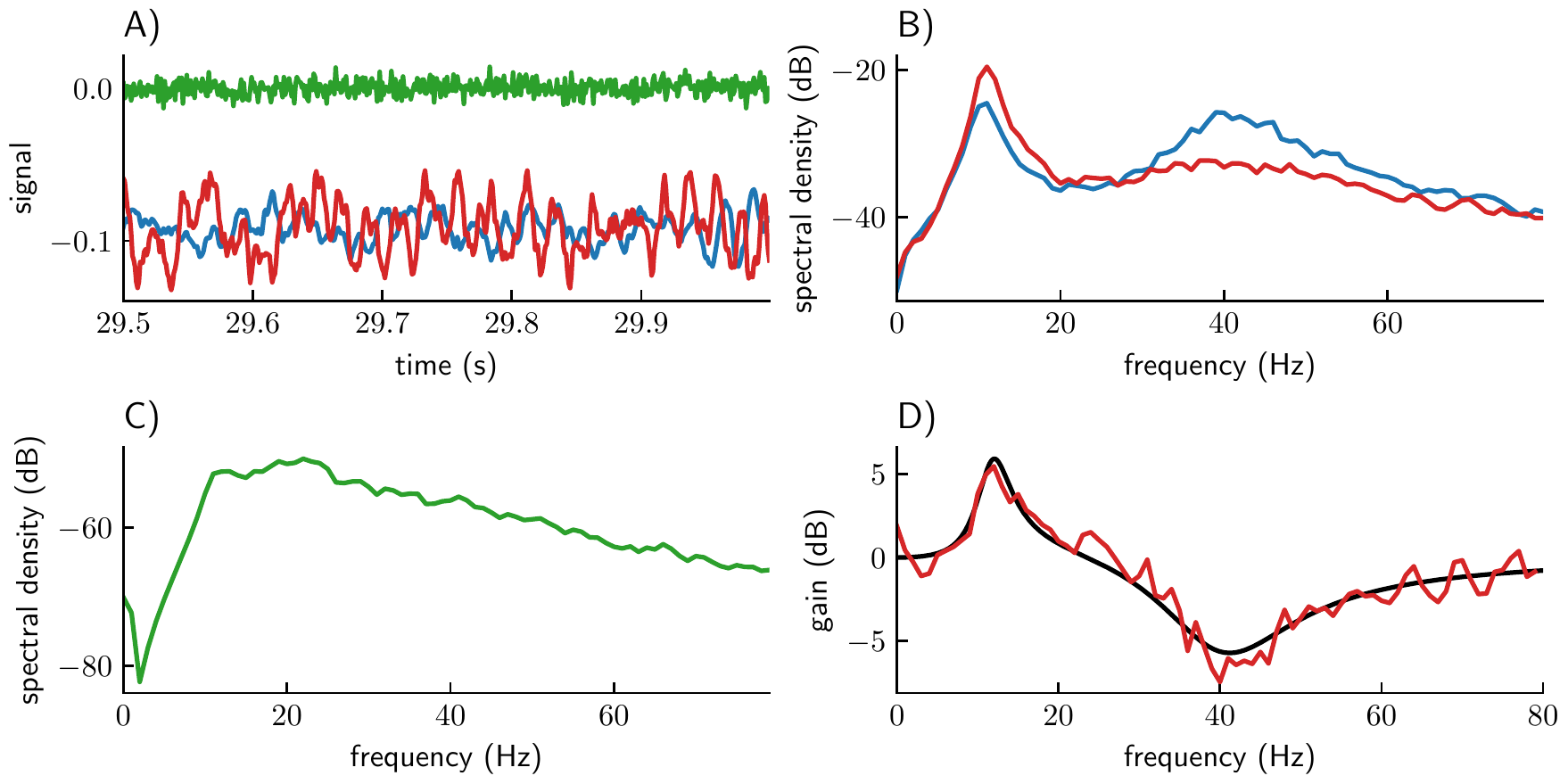}
    \caption{
        {\bf Closed-loop neurostimulation based on the estimated brain model successfully increases \textalpha-activity and decreases \textgamma-activity of the EEG signal.}
        A) Time series of the resting state activity (blue), the stimulated brain activity (red) and the stimulation current (green). B) Spectral densities of the output signals computed from their time series. C) Spectral densities of the input signal computed from the time series. D) Closed-loop gain computed from the data (red) compared to the desired closed-loop transfer function $1 + H(s)$ (black), cf. Eq.~(\ref{eq:filter}).
    }
    \label{fig:result}
\end{figure}

Simulation results show that the proposed control method is able to successfully apply the desired modifications to the EEG frequency distribution, see Fig.~\ref{fig:result}. Our results highlight that the proposed closed-loop neurostimulation method was effective in modifying the EEG frequency distribution. The simulation study used an estimated non-parametric brain model to control the EEG signal, and the results demonstrate increased \textalpha-activity and decreased \textgamma-activity. The comparison of the closed-loop gain obtained from the data to the desired transfer function further confirms the success of the proposed method in implementing the desired modifications in the EEG signal.

\begin{figure}
    \centering
    \includegraphics[width=\linewidth]{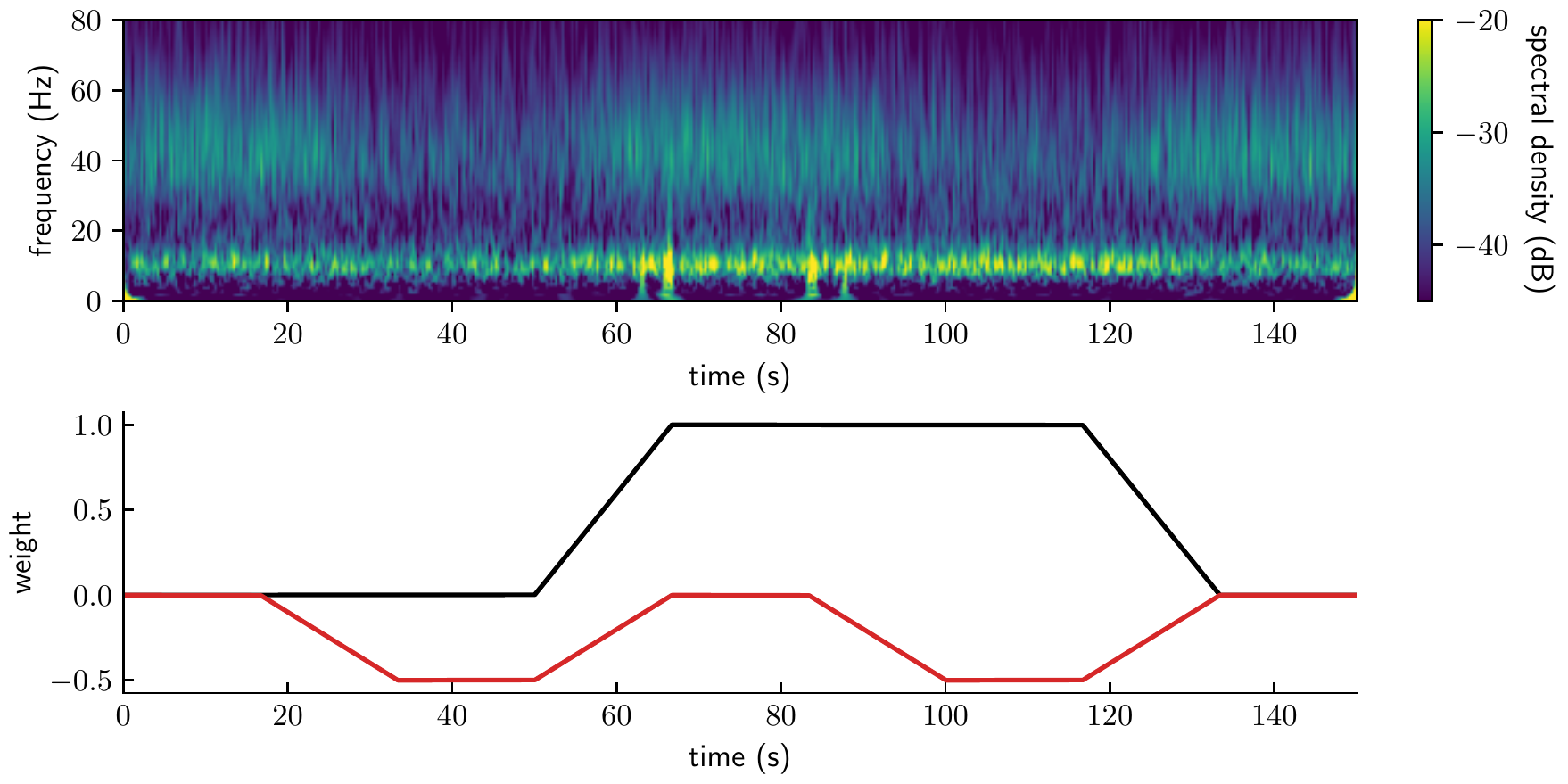}
    \caption{
        {\bf Adjusting the weights of the closed-loop filter allows      tuning  the frequency distribution of the EEG signal in real time.}
        The upper panel presents the time-frequency spectral power distribution of the simulated EEG signal utilizing the wavelet transform. It demonstrates the real-time tuning of the frequency distribution through adjustments in the weights of the closed-loop filter. The lower panel shows the evolution of these weights over time. The black curve corresponds to the weight $c_1$ of the \textalpha-band of the filter, and the red curve corresponds to the weight $c_2$ of the \textgamma-band, cf. Eq.~(\ref{eq:filter}).
    }
    \label{fig:time-frequency}
\end{figure}

These results highlight the potential of the proposed closed-loop neurostimulation in EEG-based applications and provide evidences for its feasibility in modifying the EEG frequency distribution according to the user-defined objective. To further demonstrate the power of the proposed method, Fig.~\ref{fig:time-frequency} shows how the user may change the target frequency distribution over time and how the control loop tunes the system's activity accordingly. Here, the weight $c_1$ for the \textalpha-band and $c_2$ for the \textgamma-band are modified to enhance and/or diminish the power of the corresponding system's activity. In medical neurostimulation practice, this real-time tuning of the frequency distribution permits the medical personnel to adapt the brain activity according to the patient's need. 

\section{CONCLUSIONS}\label{sec:concl}

The proposed closed-loop feedback control scheme permits to tune the frequency distribution of the brain observations (EEG) according to the users desired frequency distribution. The method was proved successful on a non-linear brain model, accordingly to our assumption of linearity of the response to small input signals. Our results highlight the potential of the closed-loop filter in tuning the frequency distribution of EEG signals in real-time. This real-time tuning of the frequency distribution may open new avenues for developing brain-computer interfaces and other EEG-based applications. Further research will be needed to test this method in a real experimental setup, however, its adaptability to each patients, and the ability to reliably tune the activity in chosen frequency domains already makes it a promising starting point for future research in closed-loop neurostimulation in clinical practice.

\addtolength{\textheight}{-12cm}   % This command serves to balance the column lengths

%%%%%%%%%%%%%%%%%%%%%%%%%%%%%%%%%%%%%%%%%%%%%%%%%%%%%%%%%%%%%%%%%%%%%%%%%%%%%%%%

%%%%%%%%%%%%%%%%%%%%%%%%%%%%%%%%%%%%%%%%%%%%%%%%%%%%%%%%%%%%%%%%%%%%%%%%%%%%%%%%

%%%%%%%%%%%%%%%%%%%%%%%%%%%%%%%%%%%%%%%%%%%%%%%%%%%%%%%%%%%%%%%%%%%%%%%%%%%%%%%%
%\section*{APPENDIX}

\section*{ACKNOWLEDGMENT}

%\color{gray}
The authors acknowledge insightful discussions with J. Riedinger.

\section*{FUNDING}
This research was funded by Inria in the "Action Exploratoire" project \em A/D Drugs.

%%%%%%%%%%%%%%%%%%%%%%%%%%%%%%%%%%%%%%%%%%%%%%%%%%%%%%%%%%%%%%%%%%%%%%%%%%%%%%%%

\bibliographystyle{unsrt}
\bibliography{main}

%\begin{thebibliography}{99}
%\end{thebibliography}

\end{document}